# Two-dimensional adaptive membranes with programmable water and ionic channels

Daria V. Andreeva[1,2], Maxim Trushin[1], Anna Nikitina[1,3], Mariana C. F. Costa[1,2], Pavel V. Cherepanov[1,7], Matthew Holwill[4], Siyu Chen[2], Kou Yang[2], See Wee Chee[5,6], Utkur Mirsaidov[1,2,5,6], Antonio H. Castro Neto[1,2], Kostya S. Novoselov[1,2,4,8]

[1]Centre for Advanced 2D Materials, National University of Singapore, 117546 Singapore

[2]Materials Science and Engineering, National University of Singapore, 117575 Singapore

[3]ITMO University, St. Petersburg 191002, Russian Federation

[4]School of Physics and Astronomy, University of Manchester, Oxford Road, Manchester M13 9PL, UK

[5]Department of Physics, National University of Singapore, Singapore, 117551 Singapore

[6]Centre for BioImaging Sciences and Department of Biological Sciences, National University of Singapore, Singapore, 117557 Singapore

[7]School of Chemistry, Monash University, Victoria 3800, Australia

[8]Chongqing 2D Materials Institute, Liangjiang New Area, Chongqing, 400714, China

**Membranes are ubiquitous in nature with primary functions that include adaptive filtering and selective transport of chemical/molecular species. Being critical to cellular functions, they are also fundamental in many areas of science and technology. Of particular importance are the adaptive and programmable membranes that can change their permeability or selectivity depending on the environment. Here, we explore implementation of such biological functions in artificial membranes and demonstrate two-dimensional self-assembled heterostructures of graphene-oxide and polyamine macromolecules, forming a network of ionic channels that exhibit regulated permeability of water and monovalent ions. This permeability can be tuned by a change of pH or the presence of certain ions. Unlike traditional membranes, the regulation mechanism reported here relies on interactions between the membranes internal structure and ions. This allows fabrication of membranes with programmable, predetermined permeability and selectivity, governed by the choice of components, their conformation and charging state.**

Adaptive translocation of various chemical species is one of the most important functions of biological membranes[1]. Among the chemicals where permeability is managed by such membranes, water and ions are the most important – their regulated transport is essential for most biological processes including synthesis, molecular degradation and recycling inside various cell compartments[2]. In most biological cells, water and ionic flow through cellular membranes is controlled by the osmotic pressure that arises due to the concentration gradient of certain ions created by regulatory proteins. Such proteins specifically interact with ions and change conformation depending on their environment[3]. Adaptation capability of biological membranes is genetically programmed[4].

Artificial adaptive systems with programmable properties, simulating the behavior of living systems, are of great interest for modern technology such as extraction of $Li^+$ for Li-ion batteries[5] and $Cs^+$ removal from radioactive waste[6]. Membranes and ionic pumps that can change their permeability and capacity depending on external conditions (such as pH or ionic concentration), mimicking biological functions, would strongly benefit fundamental and applied science. Artificial systems that can regulate water and ionic transport depending on the external parameters are in demand for adaptive filtration, self-powered bionics, and low energy neuromorphic devices *etc*.[7] Selective and dynamic regulation of water and ionic flow through artificial membranes is not currently possible: in the vast majority of modern membranes it is determined simply by their porous structure and the gradient of osmotic pressure across them[8-12].

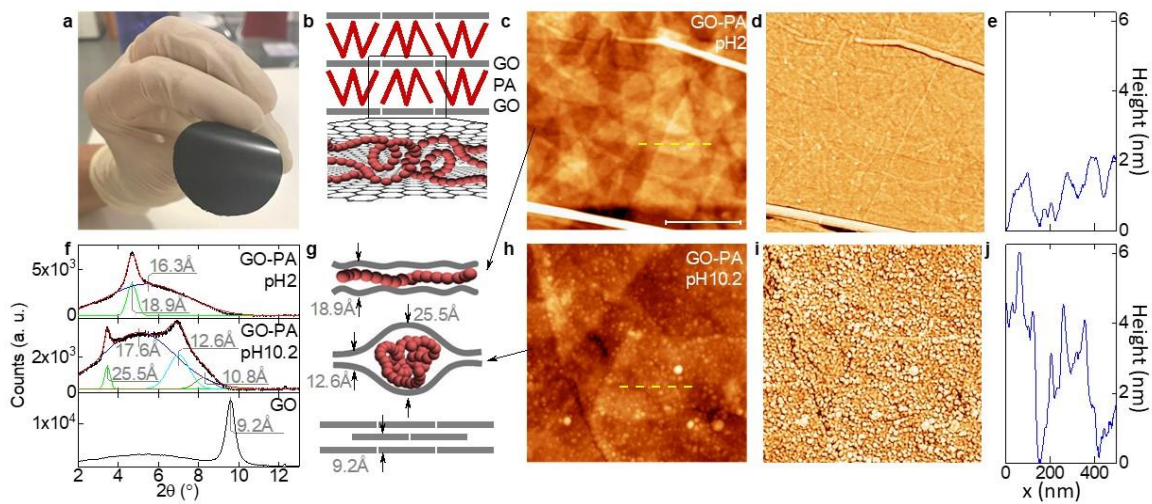

**Fig. 1. Structure of GO-PA membranes.** (**a**) Photograph of a 1-μm-thick free-standing GO-PA 25 kDa membrane. (**b**) Schematic illustration of the sandwiched architecture of our membranes. Grey lines – GO layers, red zig-zag lines – PA molecules. (**c,d**) AFM height (c) and phase (d) images of the GO flakes assembled with PA at pH2. (**e**) Height profile taken along the yellow dashed line in (c). (**f**) X-ray reflection spectra for the GO-PA membranes prepared at pH2 (top), at pH 10.2 (middle) and the GO without PA (bottom) with extracted inter-layer distances. (**g**) Schematic illustrations for the membranes' structure depending on pH of the solution during assembly based on the XRD data shown in (f). (**h,i**) AFM height (h) and phase (i) images of the GO flakes assembled with PA at pH 10.2. (**j**) Height profile taken along the yellow dashed line in (h). Scale for (c,d,h,i) is given by the scale bar in (c) which is 0.5 μm. Z-scale for (c) and (h) from black to white is 0 to 20 nm. Z-scale for (d) and (i) from black to white is 0 to 60 degrees.

Here, we propose new van der Waals composite membranes whose water and ionic permeability can be controlled by the protonation state of the components and the presence of specific ions. The structure of our membranes with adaptive permeability is analogous to rigid cellular plant membranes[13-15] with the function of the ionic channels played by the functional groups of polyamine (PA) macromolecules and graphene-oxide (GO). The rigid walls[16-18] and strong interactions in the van der Waals structure prevent swelling of hydrophilic components and provide structural stability to the membranes. Unlike typical membranes[19-21] where water and ionic flow are controlled by pore size and the external pressure, the permeability through our membranes is determined solely by the charging (protonation) state of the components of the van der Waals structure and specific interactions with certain ions. Thus, water permeability is determined by the intrinsic osmotic pressure within the interior of the rigid membrane, which arises due to excessive proton concentration due to specific (competing) charging states of GO and PA. Similarly, the ionic transport is controlled by the competing interaction between protons and ions with the functional groups at the GO-PA interfaces, which has strong selectivity towards ions, thus allowing selective ionic pumping. Furthermore, we observe that permeability of some ions can be regulated by the presence of other ions, creating a "transistor effect" for selective ionic transport.

The ordered membrane structure was formed by self-assembly of GO and PA macromolecules, Fig. 1a-b. The preparation procedure is described in the Methods. Briefly, a suspension of GO (average size of flakes - 2μm) was mixed with polyethyleneimine for 10 min. We used low (25 kDa) and high (750 kDa) molecular weight branched polyethyleneimine. Due to the opposite charges on the GO and PA molecules and low entropy of macromolecules, GO flakes become completely covered by the polymer, as confirmed by AFM images in Fig. 1c-e and h-j. After thorough washing (which ensures that only one monolayer of PA molecules is adsorbed on GO flakes) and a slow reduction of a volume of solvent via vacuum filtration - the suspension of PA decorated GO particles is collapsed in a highly ordered composite membrane with a regular distribution of monolayers of PA and GO. The measurements of the surface potential of the membranes show that the outermost layer is formed by negatively charged GO (Supplementary Fig. S1).

Polyethyleneimine is a weak polycation and its degree of ionization strongly depends on pH[22,23]. The degree of ionization of GO is also pH dependent, where it can be considered as a weak polyanion[24]. For control purposes, we also prepared bare GO membranes and membranes where GO is covered with poly(diallyldimethylammonium) chloride (PDADMAC), which is a positively charged strong polyelectrolyte and its dissociation doesn't depend on pH[25]. The thermogravimetry analysis (TGA) shows that the mass ratio of GO to PA (25 kDa) in our membranes is approx. 1:1 (Supplementary Fig. S2), and the differential scanning colorimetry (DSC) (Supplementary Fig. S2) curve shows no step typical for a glass transition of PA at -57°C. Thus, we can assume that PA forms highly mobile monomolecular layers confined in the rigid van der Waals structure. Due to self-assembly nature of the composites (PA is first assembled as a monolayer on the surface of GO, due to electrostatic interactions, and then packed together to form laminated membrane structure) the ratio between GO and PA of a particular molecular weight in our membranes is fixed.

The typical thickness of our membranes used in permeability experiments is 250 nm for GO-PA 25 kDa and 500 nm for GO-PA 750 kDa (Supplementary Figs. S2 and S3) and can be regulated by the concentration of suspension used for vacuum filtration and the molecular weight of polymer (Supplementary Fig. S3). The XRD and AFM data confirm (Fig. 1c-j) that the membranes obtained by this method indeed have a layered structure of alternating GO and PA monolayers. Note that the conformation of PA macromolecules depends on pH of the suspension during the assembly. Depending on the pH of the suspension, it is possible to deposit PA in a form of coils (Fig. 1h,i) or stretched chains (Fig. 1c,d) and to obtain two different structures as illustrated in Fig. 1g. The XRD measurements show that after assembly, the interlayer distance is insensitive to the humidity conditions (Supplementary Fig. S4).

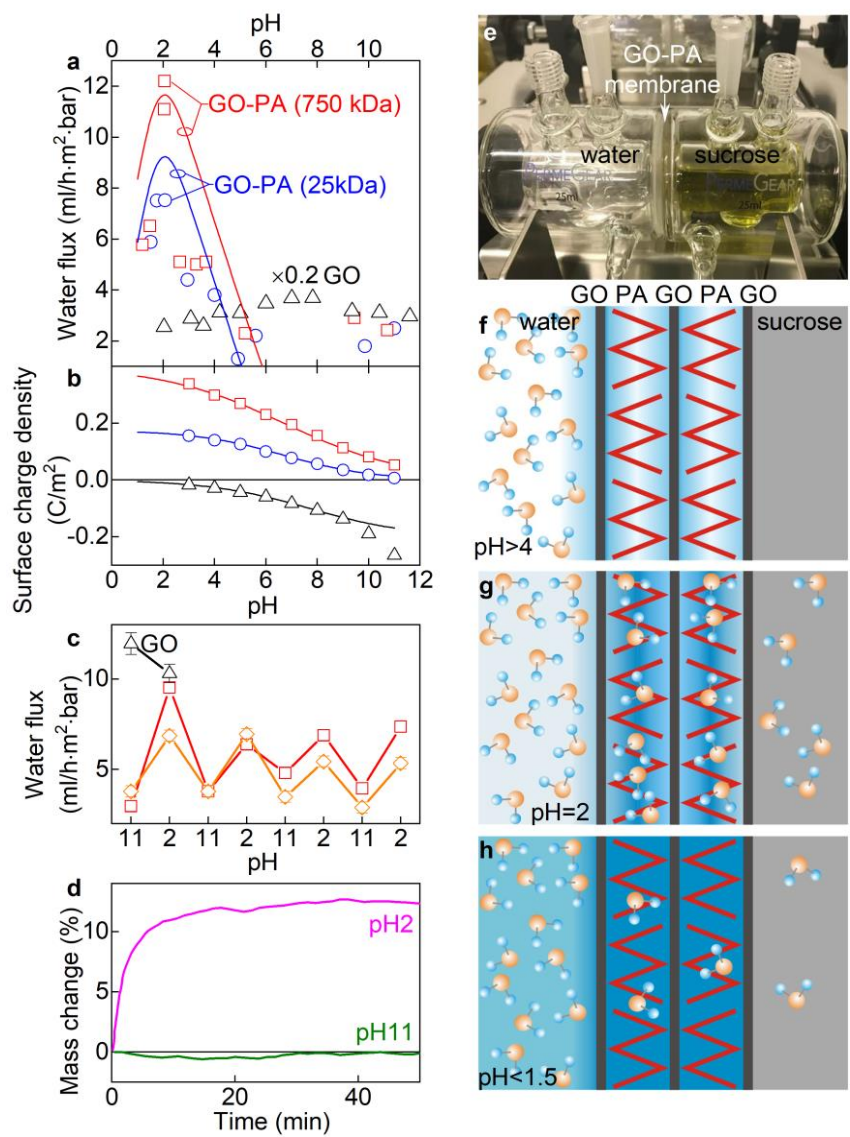

**Fig. 2. Selective permeability of GO-PA membranes.** (**a**) Water flux as a function of pH for two of our GO-PA membranes (red squares – for PA with molecular weight 750 kDa, blue circles – for PA with molecular weight 25 kDa). We tested 10 membranes. The measured water flux is fitted by the formula $c\ln(\rho_{in}/\rho_{out})$, where $c = 3.2$ (ml/h·m²·bar) is the only fitting parameter. Here, $\rho_{in}$ and $\rho_{out}$ are the H⁺ concentrations between GO layers and near the outer wall of the membrane respectively (see the main text). (**b**) Dependence of surface charge density on pH for pristine GO flakes (black triangles), GO-PA flakes with PA of molecular weight 750 kDa (red squares), and GO-PA flakes with PA of molecular weight 25 kDa (blue circles). The measurements are fitted by the Henderson-Hasselbalch equation with a = 4.5, $\sigma_0 = 0.2$ C/m² for GO, a = 4, $\sigma_0 = 0.175$ C/m² for GO-PA (25 kDa), a = 5.5, $\sigma_0 = 0.4$ C/m² for GO-PA (750 kDa). (**c**) Water permeability through GO (black triangles) and GO-PA (red squares for membrane assembled at pH2 and orange diamonds for membrane assembled at pH10.2). Each point was measured for 24 hours. GO membrane decomposed after 24 hours of measurements. The membranes were tested under osmosis conditions against 2.5M sucrose solution. The feed solution was water adjusted to pH2 and pH11. (**d**) Mass change of GO-PA membrane in acidic (pH2, magenta curve) and basic (pH11, green curve) media. The same membrane was used for both measurements. (**e**) Photograph of a dual-cell for osmosis measurements. Right volume is filled by yellow sucrose solution for better visibility. (**f-h**) Schematic illustrations of the permeability of the membranes for water regulated by proton gradients. Proton concentration is represented by the blue color.

We measured the water flux through our membranes by using them as a separator between two reservoirs – one with water (where we can control the pH by adding HCl or NaOH) and one with 2.5M sucrose (standard osmosis experimental method[26]), Fig. 2a,e. 2.5M sucrose creates a large permanent external osmotic pressure (61 Bar) across the membrane[27] and accelerates the permeability. The water flux across the membrane as a function of pH is presented in Fig. 2a. Our GO-PA membranes are very permeable for acidic solutions and significantly less permeable for basic ones (Fig. 2f-h). The bare GO membranes behave in the opposite way compared to the GO-PA membranes: they are more permeable at basic pH and less permeable at acidic pH (though the change is significantly smaller, Fig. 2c), and the GO-PDADMAC (strong polyelectrolyte) membranes show similar permeability to water as GO membranes - that is practically independent of pH (Supplementary Fig. S5).

To check the stability of our GO-PA and reproducibility of the measurements, in a separate experiment we changed pH abruptly and measured the water flow for several hours (each measurement at a particular pH takes 24 hours) (Fig. 2c). One can see that such membranes are very stable (dramatically more than pure GO membranes), can survive multiple cycles of pH changing and can strongly adjust their permeability according to pH. Note that bare GO membranes of similar thickness decomposed after one cycle (the low stability of bare GO membranes is well documented in literature[28]). We also see that water flux depends on the conformation of PA inside of the membranes. The membranes prepared at pH2 (Fig. 1c-e) are somewhat more permeable for water than the membranes prepared at pH10.2 (Fig. 1h-j). This might be explained by hopping paths for ions along the interfaces formed by coiled (pH 10.2) and uncoiled (pH 2) polymer chains as well as the increased hydrophobic nature (Supplementary Fig. S6) of weakly charged PA coils compared to stronger charged stretched chains[29]. Thus, permeability of water through our GO-PA membranes can be reliably regulated by external pH.

To understand the membrane's water permeability mechanism, we first measured the surface charge density of GO-PA and GO flakes using the potentiometric titration method[30] and fitted the data by the Henderson-Hasselbalch equation[31] $\sigma_{\text{GO-PA/GO}} = \pm \sigma_0/(1 + 10^{(\pm \text{pH} \mp \text{pK}_a)/a})$ with the saturated surface charge density $\sigma_0$ of the order of 0.1 C/m² (Fig. 2b). The fitting suggests that the charge density reaches half-saturation at pH = $\text{pK}_a$ with $\text{pK}_a$ =7.7 for pristine GO and $\text{pK}_a$ =6.7 for GO-PA. The remaining parameters are given in the caption of Fig. 2b. GO-PA flakes are strongly charged at acidic pH due to protonation of PA chains. This charge results in the excessive ionic ($H_3O^+$) concentration between GO planes where PA resides. We estimate this inner $H_3O^+$ concentration as $\rho_{\text{in}} = \sigma_{\text{GO-PA}}/ed$, where $e$ is the elementary charge, $d \approx 2$ nm is the interlayer distance, and $\sigma_{\text{GO-PA}}$ is shown in Fig. 2b for the two used PAs.

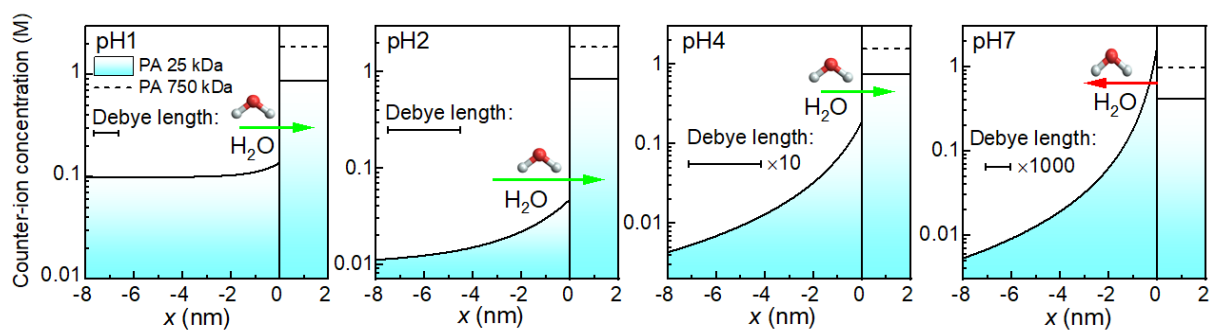

**Fig. 3. The inner and outer counter-ion concentrations creating osmotic water flow, as shown by arrows at pH1, pH2, pH4 and pH7.** We use a model structure for our composite membrane with two GO layers being placed at x=0 and x=2nm and PA occupying the space in between. The water solution occupies semivolume at x<0, and the permeation chamber with sucrose occupies semivolume x>2nm (not shown). The Debye length determines the outer concentration dependence according to the Poisson-Boltzmann equation with the GO surface charge density as an input parameter.

GO flakes form the outer walls of the membrane and are therefore exposed to the environment. GO planes acquire a negative surface charge in water (Fig. 2b), and the outer GO plane determines the beginning of the diffuse layer in electrokinetic theory[32]. The counter-ion concentration near the negatively charged plane is given by the Boltzmann distribution $\rho_{\text{out}} = \rho_{\text{out}}^{\infty} e^{e\phi_0/k_B T}$ and exceeds the bulk value $\rho_{\text{out}}^{\infty} \approx N_A 10^{-\text{pH}}$ (dm$^{-3}$) determined at infinity (far from the charged surface). Here, $T = 300$ K is the temperature, $k_B$ is the Boltzmann constant, $N_A$ is the Avogadro number, and $\phi_0$ is the electrostatic potential at the surface. The latter obeys the Grahame equation derived from the total surface/solution electroneutrality condition[33] and given by

$$\frac{\sigma_{\text{GO}}^2}{8\epsilon_0 \epsilon k_B T \rho_{\text{out}}^{\infty}} = \sinh^2\left(\frac{e\phi_0}{2k_B T}\right), \qquad (1)$$

where $\epsilon_0$ is the dielectric constant, $\epsilon \approx 80$ is the water relative dielectric permeability, and $\sigma_{\text{GO}}$ is the GO surface charge shown in Fig. 2b by the black curve. As illustrated in Fig. 3 the inner counter-ion concentration (0nm<x<2nm) is assumed to be homogeneous (determined by the protonation state of the PA), whereas the outer one depends on the distance from the membrane's wall in accordance with the Poisson-Boltzmann equation. The outer counter-ion concentration acquires its maximal value $\rho_{\text{out}}$ near the outer wall and then decreases into the bulk of water, approaching $\rho_{\text{out}}^{\infty}$. The inner excess concentration $\rho_{\text{in}} - \rho_{\text{out}}$ for PA with 25 kDa reaches a maximum at pH≈2. At higher pH the difference between inner and outer concentrations gradually vanishes and eventually becomes negative.

The known inner and outer counter-ion concentrations allow us to estimate the chemical potential difference $\Delta\mu = k_B T \ln(\rho_{\text{in}}/\rho_{\text{out}})$, which has a maximum of about 100 meV at pH≈2, see Supplementary Fig. S7. The potential difference creates an excess osmotic pressure between the diffusion layer and interior of the membrane[34]. It is further tunable by adjusting pH of the adjacent solution. The water flux is assumed to be proportional to $\Delta\mu$ resulting in controllable water transport across the membrane (Fig. 2a,c).

The model suggests that water accumulates inside the membrane at pH≈2. To confirm this, we measured water uptake by the membrane using a quartz-crystal microbalance (QCM). We found that the same membrane uptakes water at acidic pH but it does not at basic pH (Fig. 2d). The measurements on the same membrane can be repeated multiple times with very good reproducibility (Supplementary Fig. S8) by drying it in air and placing it back into water solution.

It is possible that the intake of water due to excess osmotic pressure at low pH promotes swelling of the membrane in the vertical direction and contraction in the lateral due to the Poisson effect. The in-situ liquid cell transmission electron microscopy experiments show that the wrinkled GO flakes in our membranes are unfolded in the presence of water (Supplementary Fig. S9), showing adaptivity of our membranes and forming potential for self-healing applications.

According to the QCM data (Supplementary Fig. S8), the maximum quantity of water that the membranes can absorb is approx. 12 wt.% at low pH, which corresponds to the presence of $2\times10^{15}$ molecules of water per layer. A water molecule measures approximately 0.3 nm thus covering an area $1.8\times10^{-4}$ m$^2$, which corresponds approximately to the area of our membrane. Hence, $2\times10^{15}$ of water molecules form just a single monolayer per unit cell in the most hydrated state. This confirms again that the membrane retains a perfectly integrated layered structure even when it is soaked in water.

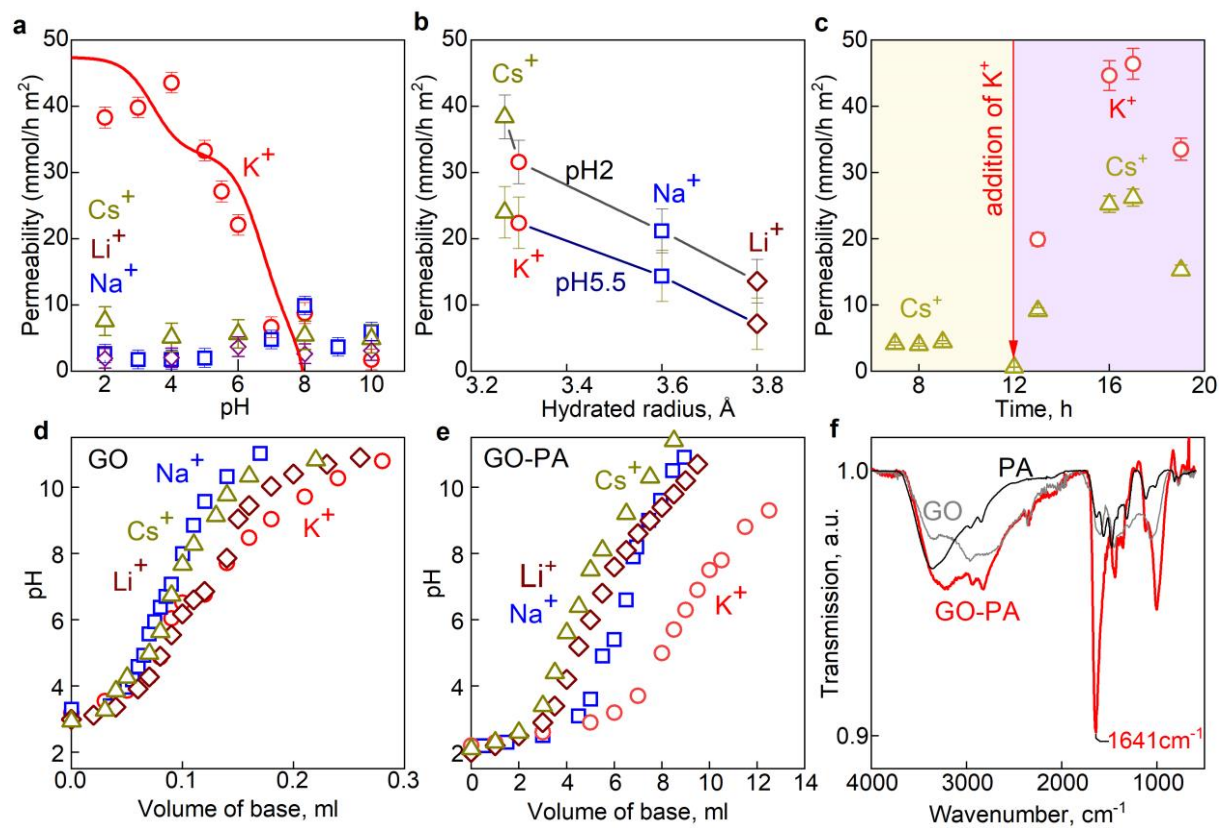

Figure 4: **pH regulated ion permeability through GO-PA 25 kDa membrane.** (**a**) Ionic permeability of GO-PA membranes as a function of pH measured separately for K$^+$, Na$^+$, Cs$^+$, Li$^+$ from 0.1M aqueous solutions of corresponding salts. Red line – a fitting curve deduced from the combinatorial model explained in Supplementary Information; (**b**) Ionic permeability vs. the hydrated radii of ions under osmosis conditions against 2.5M sucrose solution for the 0.1M mixture of ions adjusted to pH 2 and pH 5.5. We used 0.1M solutions of corresponding chloride salts adjusted to a particular pH. Overall 3 membranes were tested for each pH, and statistical distribution is given by the error bars. The experimental time was 24h. (**c**) Ionic permeability of GO-PA membranes for Cs$^+$ only for 12h, after 12h K$^+$ were added and permeability was measured for both K$^+$ and Cs$^+$ for the mixed 0.1M salt solution at pH5.5; (**d,e**) Potentiometric titration curves for suspensions of GO and GO-PA by 0.1M base (NaOH, KOH, CsOH and LiOH). Titrations started from acidic solutions prepared by adding 0.1M HCl; (**f**) ATR-FTIR spectra of the GO, PA and GO-PA membranes prepared at pH 10.2.

We also studied ionic permeability through the membranes. To this end, we used the same setup as before (Fig. 2e), where the membranes separate two volumes: one with aqueous solution of monovalent cations (here we control the pH and concentration of the four ions Cs$^+$, K$^+$, Na$^+$, Li$^+$) and another one with the 2.5M solution of sucrose. Note that the GO-PA 750 kDa membranes were completely impermeable for any of the studied ions (Supplementary Fig. S10), probably due to a large positive charge density of PA and thus, electrostatic rejection of cations typical for polyelectrolytes[35,36]. In addition, such low permeability of the membranes prepared with a high molecular weight polyelectrolyte can also be related to longer hopping paths for ions along long polymer chains. In contrast, GO-PA 25 kDa membranes demonstrate measurable permeability. Thus, we can tune the ionic permeability of our membranes without sacrificing the water permeability.

Two sets of experiments were performed on GO-PA 25 kDa membranes. In the first one, we used only one particular salt in solution (0.1M solutions of either KCl, NaCl, CsCl and LiCl) and measured ionic permeability for different pH (Fig. 4a). Concentration of ions was measured by inductively coupled plasma–optical emission spectroscopy (ICP-OES) analysis 24 hours after the permeability test. We found that $K^+$ permeability through the GO-PA 25 kDa membrane is a strong function of the pH, whereas the permeability of $Na^+$, $Cs^+$ and $Li^+$ remains low (20 times lower than that of $K^+$) and is almost independent of the pH, Fig. 4a. The transport of $K^+$ (Fig. 4a) and water demonstrates similar dependence on pH (Fig. 2a). In general, the fast permeability of small ions in pure GO (Supplementary Fig. S11) is explained in terms of capillary pressure[37-41].

In the second experiment we tested the permeability of a mixture of 0.1M solutions of the four chloride salts at pH 2 and 5.5 (Fig. 4b). Surprisingly, when in the mixture, we observe a finite permeability for all four ions (not only for $K^+$), which also depends on the hydrated radius of the ion and on the pH: ions with low hydrated radius permeate more intensively than those with larger radius and the permeability is increased at low pH. We conjecture that the presence of $K^+$ stimulates transport of other ions as well. To test this hypothesis, we performed the permeability experiment of $Cs^+$ first on its own and then in the presence of $K^+$. As in the previous experiments, $Cs^+$ doesn't permeate when on its own in solution, Fig. 4c. After 12 hours, upon adding 0.1M $K^+$ to the solution (Cs to K ratio is 1:1) we started to observe finite concentration of both ions in the drain side. Thus, the system behaves as a chemical transistor, adding the signalling component ($K^+$) opens flux of the main compound ($Cs^+$).

Selective permeability of ions with different hydrated radius has been reported previously[8] as well as non-selective electrostatic rejection of positively charged ions by polycations[35]. The selectivity of 2D membranes was dependent on the hydrated ionic radii with the cut-off around 4.5Å, which is clearly not the case in the GO-PA membranes: $Cs^+$ with smaller hydrated radius permeates slower than $K^+$, and the radius of both is around 3.25Å. Furthermore, interlayer spacing in our membranes is significantly larger (1.9-2.5 nm) than in bare GO membranes (0.9-0.12 nm)[35,42].

To explain the significant difference in permeability of $K^+$ and other ions as well as the cumulative effect of ion permeability, we studied interaction of GO and GO-PA with ions using potentiometric titration by 0.1M NaOH, KOH, CsOH and LiOH (Fig. 4d,e). There is a very small difference for titration of GO in the four different bases, Fig. 4d. However, we observe a strong shift in measured pK for the titration of GO-PA by KOH, Fig. 4e. The titration curve for GO-PA by KOH indicates that a large number of functional groups in GO-PA are accessible for $K^+$. $K^+$ replaces protons in such groups, releasing them to the medium, thus shifting the measured pK. Indeed, ICP analysis of the membranes immersed in salt solutions show that the concentration of $K^+$ inside of the membrane is significantly higher when the concentrations of $Na^+$, $Cs^+$ and even $Li^+$, see Supplementary Fig. S12.

Currently we do have several explanations why our GO-PA membranes have such a strong affinity to $K^+$. In nature, selectivity of potassium channels in biological membranes is explained by specific interactions[43] of carbonyl oxygen atoms with $K^+$. Coordination of carbonyls is also responsible for transport of $K^+$ by a $K^+$ selective ionophore, valinomycin[44]. $Na^+$ and $Li^+$ cannot be transported because the dehydrated ions of Na and Li are too small to interact with the inward-facing carbonyls of peptides[45]. To this end we use X-ray photo electron spectra (XPS) and attenuated total reflectance Fourier transform infrared spectra (ATR-FTIR) (Fig. 4f and Supplementary Fig. S13) to monitor the presence of carbonyls in the GO-PA membrane. Whereas the carbonyls' corresponding peak is very weak in pure GO (Fig. 4f) – such peak appears strongly in GO-PA membranes, which suggests that due to electrostatic interactions between GO and PA – carbonyl groups are exposed to their interface. Due to the specific affinity of $K^+$ to carbonyl groups at the interfaces formed by GO and PA the entire membrane acts as an ionic pump. Once the affinity is established (like in the case of GO-PA and $K^+$) the ion hopping transport is possible along the interfaces that form the channels.

Based on these observations we suggest the following model for the ionic transport through GO-PA membranes. In the normal state GO-PA membranes are closed for the transport of hydrated ions due to strong repulsion between the cations and the positively charged PA. However, the chain of carbonyl groups at the interface of GO and PA act as selective ionic channels, allowing dehydrated $K^+$ to permeate through. Large $K^+$ ions might even replace protons on GO-PA interfaces, reducing the overall charge, thus opening channels for hydrated ions of any species. Thus $K^+$ acts as a sort of a gate in an "ionic transistor". Furthermore, the sensitivity of such a "transistor" to the presence of $K^+$ ions can be regulated by the overall protonation state of PA – through the external pH.

The proposed model is highly phenomenological and thus it is very difficult to provide a quantitative description for it. The key mechanism that opens the ionic channels for the transport of hydrated ions is the replacement of a fraction of protons in the interlayer PA with $K^+$ ions. The $K^+/H^+$ exchange can be treated as a stochastic process of competition between protons and potassium ions for the available sites on GO-PA. A simple combinatorial model suggests that the maximum of the $K^+/H^+$ exchange frequency should be achieved when the surface densities of $H^+$ and $K^+$ inside the membrane are approximately the same. The result of this model presented in Fig. 4a fits the dependence of potassium permeability as a function of pH reasonably well. The model also suggests that permeability for $K^+$ is always higher than for $Cs^+$ in the $K^+/Cs^+$ mixture because $K^+$ can exchange with both $H^+$ and $Cs^+$ whereas $Cs^+$ can do that with $K^+$ only. This is indeed observed in Fig. 4c: the permeability ratio for $K^+$ and $Cs^+$ is larger than 1 and does not depend on time. Similar effect can also be seen in permeability for $Na^+$ and $K^+$ (Supplementary Fig. S14). The model based on the fitted surface charge densities (Supplementary Fig. S15 and Table S1) is further illustrated in Supplementary Figs. S16 and S17.

The biomimetic behavior of our membranes in terms of controlled water transport and selective ionic permeability is a striking feature that is essential for the creation of adaptive membranes. Furthermore, our membranes are operational and highly dynamic even in physiological conditions. The ionic selectivity and permeability can be further programmed during the preparation stage (for instance by the choice of strength of the polyions and their molecular weight), opening multiple opportunities for applications. For instance, it should be possible to design membranes with regulated selective $K^+/Na^+$ pumping, for the extraction of $Li^+$ or separation of $Cs^+$. We also stress that our membranes use a very novel mechanism of regulated water and ion permeability. Gaining further control on such mechanisms will be of fundamental importance for understanding and construction of stimuli responsive composite materials. Such membranes will allow building relatively simple artificial structures that reproduce such properties of living matter as switchable ionic permeability and selectivity and will lead to further advances in the formation of super-nanocapacitors, membranes with selective release of ions for biofilm growth, membranes for extraction of $Li^+$ for lithium ion batteries and extraction of radioactive $Cs^+$ for water purification as well as for iontronics and neuromorphic devices. Thus, membranes operating on the principles we discussed can be the key components for the construction of artificial membranes with intrinsic intelligence.

## Methods

The membranes were prepared by vacuum filtration of suspensions of graphene oxide flakes (graphene oxide, 2 mg/mL, dispersion in $H_2O$, Merck) covered by polyethyleneimine (PEI, branched average Mw ~25,000 by light scattering (LS), branched average Mn ~10,000 by gel permeation chromatography (GPC), (Merck)) and PEI (branched average Mn ~60,000 by GPC, branched average Mw ~750,000 by LS, 50 wt. % in $H_2O$ (Merck)) on two types of filters: Anodisc 47 ( pore size – 0.02 µm, diameter 47 mm, Whatman, UK) and polyethersulfon membrane (0.03 µm, 47 mm, Sterlitech Corporation, USA). Surface charge of GO and PA-GO flakes was estimated using a titration technique. The titrations of 0.5g GO or GO-PA were performed in 0.005 M NaCl background electrolyte. Titrations started either from lower pH (GO dispersion), or higher pH (GO-PEI dispersion) using 0.1 M NaOH or 0.1 M HCl as base/acid titrants. Both dispersions were titrated in the pH range from 2 to 12. Water flux was measured by using a H1C side-by-side diffusion system equipped with an H1C magnetic stirrer and a heater/circulator (PermeGear, USA). The membrane was placed between two compartments. In one half, we added 2.5 M solution of sucrose (99%. Merck) in water to create a pressure of approx. 61 Bar. In the second half of the cell, we used water solutions at particular pH that was adjusted using 1M HCl or 1M NaOH solutions. The permeability tests were conducted for 24 h. Perkin Elmer Optima 5300DV was applied to measure the concentration of ions by inductively coupled plasma - optical emission spectroscopy (ICP-OES). Morphology was studied by AFM, Dimension FastScan AFM (Bruker, USA), the cross section by SEM using Field Emission Scanning Electron Microscope (FESEM) FEI Verios460 (ThermoFisher Scientific, USA) at an operating voltage of 2 keV, structure by XRD using Bruker D8 Advance X-ray transmission diffractometer (Bruker, USA) (CuKα radiation from the copper target using an in built nickel filer, λ = 1.54056 Å), swelling by Quartz-Crystal Microbalance (QCM) measurements on AWS A20 system (BioLogic, France) using Ti/Au polished finish 5 MHz resonators.

### Measuring proton concentration

We also directly measured how many $H^+$ the GO-PA membrane can consume and accumulate from the electrolyte to show that amino-groups of PA molecules in the membranes are accessible for $H^+$ and provide their accumulation in the midplanes of sandwiches. We used a two-compartment-cell (for details see Supplementary Information) and placed our membrane between the compartments filled with water at pH 2, and water at pH 11. Monitoring pH in both compartments shows that $5 \times 10^{-5}$ mol of protons are consumed from water at pH 2 and $1.97 \times 10^{-5}$ mol and released in water at pH 11 for 24h, therefore, $3.03 \times 10^{-5}$ mol of protons are accumulated in the membrane. Knowing weight of PA,~1.5 mg per a membrane, we can calculate how many amino-groups are actually available for $H^+$ per a membrane. $3.25 \times 10^{-5}$ amino-groups are available for $H^+$ and can accumulate $3.03 \times 10^{-5}$ mol of $H^+$. Thus, these results support our hypothesis that PA is important for accumulation of electrostatically attracted $H^+$.

## Acknowledgments

We thank NRF (Singapore) for financial support through the project Medium-sized centre programme R-723-000-001-281 and RSF (Russian Federation) grant no. 19-19-00508. MT thanks Director's Senior Research Fellowship of the Centre. KSN also acknowledges support from EU Flagship Programs (Graphene CNECTICT-604391 and 2D-SIPC Quantum Technology), European Research Council Synergy Grant Hetero2D, the Royal Society, EPSRC grants EP/N010345/1, EP/P026850/1, EP/S030719/1.

AUTHOR CONTRIBUTIONS

D. V. A., M. T. and K. S. N. conceived and designed the experiments: A. N., M. C. F. C., P. V. C., M. H., K. Y., S. C., S. W. C. and U. M. performed the experiments: A. H. C. N. contributed materials/analysis tools: D. V. A., M. T. and K. S. N. co-wrote the paper. All authors discussed the results and commented on the manuscript.

DATA AVAILABILITY STATEMENT: The data that support the plots within this paper and other findings of this study are available from the corresponding authors on reasonable request.

ADDITIONAL INFORMATION

Supplementary information is available in the online version of the paper. Reprints and permission information is available online at www.nature.com/reprints. Correspondence and requests for materials should be addressed to [D. V. A., M. T. and K. S. N.]